\documentclass[prd,aps,showpacs,nofootinbib]{revtex4}
\usepackage{amssymb}
\usepackage{graphicx}
\usepackage{amsmath}

\newcommand{\be}{\begin{equation}}
\newcommand{\ee}{\end{equation}}
\newcommand{\bq}{\begin{eqnarray}}
\newcommand{\eq}{\end{eqnarray}}

\begin{document}
\title{Thompson's renormalization group method applied to QCD at high energy scale}
\author{* Cl\'audio Nassif, * J. A. Helayel-Neto and P. R. Silva}
\affiliation{\small{* CBPF-Centro Brasileiro de Pesquisas F\'{\i}sicas, Rua Dr. Xavier Sigaud,150,
CEP 22290-180, Rio de Janeiro-RJ, Brazil.\\
Departamento de F\'{\i}sica - ICEx - UFMG~
(C.P 702 - 30.123-970 - Belo Horizonte - MG - Brazil)}}

\begin{abstract}

 We use a renormalization group method to treat QCD-vacuum behavior specially closer to the regime of
 asymptotic freedom. QCD-vacuum behaves effectively like a ``paramagnetic system" of a classical theory in the sense
 that virtual color charges (gluons) emerges in it as a spin effect of a paramagnetic material when a magnetic field
 aligns their microscopic magnetic dipoles. Due to that strong classical analogy with the
paramagnetism of Landau's theory,we will be able to use a certain Landau effective action without temperature and
phase transition for just representing
QCD-vacuum  behavior at higher energies as being magnetization of a paramagnetic material in the presence of a
magnetic field $H$. This reasoning will allow us to
apply Thompson's approach to such an action in order to extract an ``effective susceptibility" ($\chi>0$) of QCD-vacuum. It depends on
logarithmic of energy scale $u$ to investigate hadronic matter. Consequently we are able to get an ``effective magnetic permeability" ($\mu>1$) of such a
``paramagnetic vacuum". Actually,as QCD-vacuum must obey Lorentz invariance,the attainment of $\mu>1$ must simply require that the ``effective
electrical permissivity" is $\epsilon<1$ in such a way that $\mu\epsilon=1$ ($c^2=1$). This leads to the anti-screening effect where the asymptotic
freedom takes place. We will also be able to extend our investigation to include both the diamagnetic fermionic
properties of QED-vacuum (screening) and the
paramagnetic bosonic properties of QCD-vacuum (anti-screening) into the same formalism by obtaining a  $\beta$-function at 1 loop,where both the bosonic and fermionic contributions are considered.
\end{abstract}

\maketitle

\section{Introduction}

In Science, there are a considerable number of problems  where fluctuations are present in all scales of length, varying from microscopic to macroscopic
 wavelengths. For instance, we can mention the problems of fully developed turbulent fluid flow, the growing of polymer chains, critical phenomena and
 elementary particle physics. The problem of non-classical reaction rates (diffusion limited chemical reactions) turns out also to be in this category.

As was pointed out by Wilson\cite{1}: ``in quantum field theory,``elementary" particles like electrons, photons, protons and neutrons turn out to have composite internal structure on all sizes scales down to zero. At least this is the prediction of quantum field theory".

The most largely employed strategy for dealing with problems involving many length scales is the ``Renormalization - Group (RG) approach".
The RG has been applied to treat the critical behavior of a system undergoing second order phase transition and has been shown to be a powerful method to obtain their critical indexes\cite{2}.

In an alternative way to the RG approach, C. J. Thompson\cite{3} used a heuristic method (of the dimensions) as a means to obtaining the correlation length critical index ($\nu$), which governs the critical behavior of system in the neighborhood of its critical point. Starting from Landau-Ginzburg-Wilson hamiltonian or free energy, he got a closed form relation for $\nu(d)$ \cite{3}, where d is the spatial dimension. It is argued that the critical behavior of this $\Phi^4$-field theory is within the same class of universality as that of the Ising Model.

One of the present authors\cite{4} applied Thompson's method to study diffusion limited chemical reaction ${\bf A+A \to 0}$ (inert product). The
results obtained in that work\cite{4} agree with the exact results of Peliti\cite{5} who renormalized term by term given by the interaction diagramms
in the perturbation theory.

More recently, Nassif and Silva\cite{6} proposed an action to describe diffusion limited chemical reactions belonging to various classes of universality. This action was treated through Thompson's approach and could encompass the cases of reactions like ${\bf A+B \to 0}$ and  ${\bf A+A \to 0}$ within the same formalism. Just at the upper critical dimensions of  ${\bf A+B \to 0}$ ($d_c=4$) and  ${\bf A+A \to 0}$ ($d_c=2$) reactions, the present authors found universal logarithmic corrections for the mean field behavior.

Thompson's renormalisation group method has been applied to obtain the correlation length critical exponent of the Random Field Ising Model by Aharony, Imry and Ma\cite{7} and by one of the present authors\cite{8}. His method was also used to evaluate the correlation length critical exponent of the N-vector Model\cite{9}. Yang - Lee Edge Singularity Critical Exponents\cite{10} has been also studied by this method. In short we have been exploring the various possibilities of the Thompson's method of dimensions\cite{4}~\cite{6}~\cite{8}~\cite{9}~\cite{10}~\cite{11}~
\cite{12}~\cite{13}~\cite{14}~\cite{15}~\cite{16}~\cite{17}. As we can see, for instance, by considering these various possibilities of the method, we were able to obtain the universal logarithmic behavior for the coupling parameters of various models at their respective upper critical dimensions [4,6,8-16]. We also have shown how this method behaves when applied to $QED_4$\cite{17} and we have obtained the logarithmic behavior on scale of energy for coupling $\alpha$ (for $d=4$).

The aim of the present work is to describe firstly the QCD-vacuum behavior by considering a strong classical analogy of such vacuum with a paramagnetic material in the presence of an external magnetic field \cite{18}. To do that, we will use in section 3 a simple action in the form of that of Landau,i.e.,without temperature and phase transition, where the magnetization due to the presence of an external magnetic field $H$ is thought as a
color scalar field of virtual gluons. Such cloud of virtual gluons are induced in vacuum because,in the investigation of the internal structure of the
nucleons, higher energy scales must be also considered. By applying Thompson's method (T.M) to such an action, it will be possible to extract an ``effective electric  permissivity" $\epsilon<1$, an ``effective magnetic permeability" $\mu>1$ and also an ``effective susceptibility" $\chi>0$ which depends on
logarithmic of energy scale $u$ used to investigate the hadronic matter. Just in order to obey Lorentz invariance, we will make the simple Lorentz
condition for vacuum, that is, $\mu\epsilon=1$ ($c^2=1$)\cite{18}. Our investigation provides an analogy between the energy of the QCD-vacuum and the
corresponding energy of magnetic dipoles of a paramagnetic material being ligned up by the action of a magnetic field. Due to this fact, we will verify
in section 2 that QCD-vacuum at higher energies behaves as if it were predominantly a kind of ``color paramagnetism" for gluons with spin 1 ,that is,
the bosonic behavior of QCD-vacuum in such ``paramagnetic regime ($\mu>1$)" for higher energies supplants completely the fermionic contribution for vacuum due to ``diamagnetic regime ($\mu<1$)", and thus we will get asymptotic freedom in QCD as a consequence of this antiscreening effect (``vacuum
paramagnetism")\cite{18}.

 Actually QCD-vacuum is more complex than QED-vacuum in the sense that the first one presents a kind of competition between the two contributions,
 namely
a screening effect (``vacuum diamagnetism",$\mu<1$ or ``vacuum dielectricity",$\epsilon>1$) and an antiscreening effect (``vacuum paramagnetism",
$\mu>1$), such that antiscreening supplants
completely screening at higher energies (asymptotic freedom). On the other hand, QED-vacuum is more simple because it presents only the
contribution for the screenig effect. In section 4, taking in account these two contributions, we will get a unified vision for QCD and QED by
obtaining a general $\beta$-function where both bosonic and fermionic contributions are considered together,such that we recover QED $\beta$-function
for 1 loop as a special case when there is no gluon state ($n_g=0$) and only one flavor ($n_F = 1$), that is, we have the special case of just fermionic
vacuum with its characteristic dielectric properties ($\epsilon>1$).


\section{$QCD$-Lagrangian, color charges, gluons and the ``paramagnetism of color fields": the asymptotic freedom} 

\subsection{$QCD$ - Lagrangian} 

Quantum Chromodynamics (QCD), the modern theory of the strong interactions \cite{19}~\cite{20} is a non-abelian field theory. In 1973, Gross and
Wilczek \cite{21} and independently Politzer\cite{22} have shown that certain classes of non-abelian fields theories exhibit asymptotic freedom, a
necessary condition for a theory that could describe strong interactions. These seminal papers\cite{21}~\cite{22} opened the route to the birth
of QCD.

 In a not very accurated picture, QCD can be considered as an expanded version of QED. In QCD we have also six fermionic fields representing the
 various
quark flavors, in contraposition to a single fermionic field of QED. Besides the asymptotic freedom exhibited at the ultraviolet limit, a theory of
the strong interactions must also display quark confinement at the infrared limit.

 Whereas in QED there is just one kind of charge, QCD has three kinds of charge, labeled by ``color" (red, green and blue)\cite{19}. The color
 charges are
 conserved in all physical processes. There are also photon-like massless particles, called color gluons, that respond in appropriate ways to the
 presence
 of color charge. This mechanism has some similarity with the ways photons respond to electric charge in QED, excepted by the non-abelian
character of the theory.

 Let us write the QCD-Lagrangian density, namely:

  \begin{equation}
L=\Sigma_j \overline\psi_j(i\gamma_{\mu}D^{\mu}- m_j)\psi_j - \frac{1}{4}G^a_{\mu\nu}
G_a^{\mu\nu},
  \end{equation}

where $D^{\mu}=\partial^{\mu}+\frac{1}{2}ig\lambda_a A^{\mu}_a$ , and $G^{\mu\nu}_a=\partial^{\mu}A^{\nu}_a-\partial^{\nu}A^{\mu}_a-gf_{abc}A^{\mu}_b
A^{\nu}_c$\cite{19} 

 In (1) above, $m_j$ and $\psi_j$ are the mass and quantum field of the quark of $j^{th}$ ``flavor", and $A$ is the gluon field, being $\mu$ and
 $\nu$ the
 space-time indexes. $a$, $b$ and $c$ are color indexes. The coefficients $f$ (structure constants) and $\lambda_a$ guarantee $SU(3)$ color symmetry.
  $g$ is the coupling constant.

 The gluon part of (1) contains both a kinetic term, $L_{kin}=-\frac{1}{4}(\partial_{\mu}A_{a\nu}-\partial_{\nu}A_{a\mu})(\partial_{\mu}A_a^{\nu} -
 \partial_{\nu}A_a^{\mu})$, and an interaction term $L_{int.}= \frac{1}{2}g f_{abc}(\partial_{\mu}A_{a\nu}-\partial_{\nu}A_{a\mu})A_b^{\mu}A_c^{\nu}
-\frac{1}{4}g^2 f_{abc} f_{a^{\prime} b^{\prime} c^{\prime}} A_{b\mu}A_{c\nu} A_{b^{\prime}}^{\mu} A_{c^{\prime}}^{\nu}$. The form of the kinetic term is the same form as the photon term of the well-known QED-Lagrangian. Thus exchange of gluons gives rise to forces similar to the Coulomb interaction, but acting on particles with color instead of charges. However, gluons carry color themselves (unlike photons that don't carry any charge), which leads to the interaction term ($L_{int.}$) between gluons themselves, and this is the situation that makes QCD an asymptotically free theory 

\subsection{Color charges and color fields (gluons)} 

 It is well-known the energy stored in an electric field according to classical theory, mamely $U_{cl.}=\int_{V_3} E_{cl.}^2 dV_3$, being the
 integration performed in a $3D$ space-like volume. In a previous paper (see ref. [17]), where we have considered $QED$ at high energies, quantum
 fluctuations due to vacuum polarization affect the energy through a squared quantum contribution in $\Delta E$, since
 the linear quantum contribution term $\Delta E$ averages out to zero. So we have obtained $\overline{E^2}=\overline{E_{cl.}^2}+ \overline{\Delta
 E^2}$~\cite{17}, where the bars means averaging over a sufficiently long time in the scale of the fluctuations, and therefore we were interested in
 the quantum process namely the absorption and emission of virtual photons, leading to the quantum correction in the field $E$, that is, $\Delta
 E_{rms}=[(\overline\Delta {E^2})]^{\frac{1}{2}}$~\cite{17}, where the index $rms$ means root mean square. We have thought that such a correction is
 different from zero only in the presence of the fermionic field due to its purely quantum origin. This led us to propose the relation
$\Delta E_{rms}^2=\xi^2\psi_{rms}^2$~\cite{17}, where we have considered $\psi_{rms}^2=\left<[\overline\psi\psi]\right>_r=\frac{1}{2\pi^2 r^3}$ (see 11 in ref. [17]).  $\psi_{rms}^2$ corresponds to the mean squared fermionic field in the variable of scale-$r$, and $\xi$ is a proportionality
 constant. Such relations allowed us to obtain $\Delta E_{rms}\propto\frac{1}{r^{\frac{3}{2}}}$ in QED for quantum contribution of the field
 \cite{23}~\cite{24}~\cite{25} at high energy. It must be compared with the inverse square Gauss law of the classical contribution and so we
 perceive that it leads to a logarithmic correction on scale $r$ for energy of the field\cite{17}~\cite{25}.

 As QCD introduces color charges and color fields since gluons carry color charges (unlike photons that don't carry any charge)\cite{19}, we could
 extend the reasoning above to treat QCD by considering a general ``color electric field", namely:

 \begin{equation}
\overline{E_a^2}=\overline{E_{cl.a}^2} + \overline{\Delta E_{F.a}^2} + \overline{\Delta E_{B.a}^2}, 
 \end{equation} 
where $\overline{E_{cl.a}^2}$ is the classical contribution for the color field with a ``mode" $a$. $\overline{\Delta E_{F.a}^2}$ is a fermionic
 contribution for the color field, which is similar to that of QED ($\Delta E_{rms}=[(\overline{\Delta {E^2}})]^{\frac{1}{2}}$~\cite{17}), however
QED has no color. $\overline{\Delta E_{B.a}^2}$ is a quantum contribution for the color field, which does not have any analogy with QED. It is due to
 quantum fluctuations of color fields in the presence of bosons (gluons) since they carry color themselves, that is, it is a bosonic contribution for
 the color field. We wiil see that such a new quantum contribution is exclusively from QCD-vacuum behavior, which leads to the antiscreening
 effect and thus makes QCD an asymptotically free theory.

 Actually, relation (2) supplies a total energy density $u = u_{cl.} + u_{F} + u_{B}$, being $u_{cl}$ the classical contribution for energy density,
  $u_{F}$ and $u_{B}$ the fermionic and bosonic contributions respectively. We will see that $u_{B}$ has a changed signal with respect to $u_{F}$,
 which leads to the antiscreening effect and the asymptotic freedom in QCD, in opposition to the screening effect of QED ($u_F$).

Now we assume that a heuristic approach used by Thompson\cite{3} to study critical phenomena can be applied to the lagrangian (1). The first
 prescription of Thompson\cite{3}~\cite{17} is basically a scale argument with dimensional analysis for average values on scales. It states that:

``When we consider the integral of the Lagrangian (1) in a coherence volume $l^d$ in d-dimensions, the modulus of each integrated term of it is separately of the order of unity".

This method by using its three prescriptions was firstly applied by Thompson\cite{3} to the Landau-Ginzburg-Wilson free energy or Hamiltonian, obtaining
 critical exponents within the same universality class of the Ising model. As the present model does not have any kind of phase transistion or
 spontaneous
breakdown of symmetry, it is only necessary to use the first prescription of Thompson. So by applying such prescription to the knetic fermionic term of Lagrangian (1), we write:

\begin{equation}
\left|\int_r [\overline\psi_j(i\gamma_{\mu}\partial^{\mu})\psi_j]_r dV_4\right|\sim 1,
\end{equation} 
where $dV_4\sim r^3 dr$ 

We can observe that the dimension of \'~$\gamma_{\mu}\partial^{\mu}$\`~($[\gamma_{\mu}\partial^{\mu}]_r$) which appears in the integral (3) is the same as
 $[\partial^{\mu}]_r = r^{-1}$. This is because we are thinking only about a dimensional analysis in (3) for \'~$\gamma_{\mu}\partial^{\mu}$\`~. So in
 such case we can naturally neglect the spinorial aspect of the field and just consider the \'~first derivative $\partial^{\mu}$\`~, which defines the
 fermions (quarks) regarding to the scaling dimensional analysis,that is, $[\partial^{\mu}]_r = r^{-1}$.

It is interesting to note that the integral above leads immediately to a kind of scaling dimensional analysis, where the dimensional value of certain
quantity $[\overline{\psi_j}\psi_j]$ inside the integral is taken out of its integrand as a mean value in a coherent hyper-volume of scale $l^4$,
being $dV_4\sim r^3 dr$. Thus from (3) we extract the following scaling behavior, namely:

\begin{equation}
\left<[\overline{\psi_j}\psi_j]\right>_r \equiv [\overline{\psi_j}\psi_j]_r\sim\frac{1}{r^3}
\end{equation}

In analogous way to that heuristic reasoning used in QED by considering the fermionic contribution of condensate $\left<\overline\psi\psi\right>$ over
quantum fluctuations of field $E$, that is to say, $\overline{\Delta E^2}\propto\left<[\overline\Psi\Psi]\right>_r\propto\frac{1}{r^3}$~\cite{17}, then,
 for QCD, we have a similar fermionic contribution from quark condensate which also contributes for quantum fluctuations of the color field $E_a$ through
 $\Delta E_{F.a}$, namely:

\begin{equation} 
\overline{\Delta E_{F.a}^2}\propto\left<[\overline\Psi_j\Psi_j]\right>_r\propto\frac{1}{r^3}
\end{equation}

\subsection{``Paramagnetism of color fields" in QCD-vacuum: the asymptotic freedom}

Let us firstly recapitulate some properties of ordinary polarizable media in the classical theory. In a polarizable medium, the potential energy of
two static test charges $q$ and $Q$ is $U_{el.}(r)=\frac{qQ}{4\pi\epsilon r}$, where $r$ is the distance between the two charges, being $\epsilon$ the
dieletric constant,which in vacuo takes the value $\epsilon_0=1$. Ordinarily, the polarizability of the medium causes a screening of the interaction
between the test static charges, meaning that $\epsilon>1$. On the other hand, antiscreening corresponds to $\epsilon<1$.

A relativistic quantum field theory has a vacuum which presents a strong classical analogy with the ordinary polarizable medium, however it just differs
 from an ordinary polarizable medium on a very important aspect: it is relativistically invariant. This means that,if we set the velocity of light
 $c=1$, the magnetic permeability $\mu$ is related to the dieletric constant (electric permissivity) $\epsilon$ by

 \begin{equation}
\mu\epsilon=1.
\end{equation}

The implication of Lorentz invariance in QCD is very important in theories on confinement of quarks and gluons\cite{26}. Such a relationship (6) does not
exist for an ordinary or classical polarizable medium.

In order to obey Lorentz invariance given in (6), we can conclude that ordinary screening means $\mu<1$ (diamagnetism)
and ordinary antiscreening means $\mu>1$ (analogous to paramagnetism of the Landau's classical theory). The magnetic permeability $\mu$ is written in 
the following way:
  \begin{equation}
 \mu=1+4\pi\chi,
\end{equation}
where $\chi$ is the magnetic susceptibility. QCD-vacuum has classical analogy to the
paramagnetic medium\cite{18}. We will see that the increasing of energy scale $u$ to investigate hadronic matter leads to increasing of ``effective
susceptibility" of QCD-vacuum $\chi_{eff.}=\chi(u)(>0)$ to be determined in the next section. This leads to an increasing of the ``effective magnetic permeability" of QCD-vacuum, namely $\mu_{eff.}=1+4\pi\chi(u)$.

By considering a paramagnetic medium with a volume $V$ and an uniform magnetization $M$ in the presence of the field $H$, thus we have the following
energy:

\begin{equation}
E = E_{paramagnetic}= -\frac{1}{2} 4\pi MHV = -\frac{1}{2} 4\pi \chi H^2 V,
\end{equation}
where $M=\chi H$.

In spite of there is not Lorentz invariance in ordinary media, a paramagnetic medium still realizes a strong classical analogy to QCD-vacuum in the sense
that we could think that such a vacuum is a medium with spin effect of color charges\cite{18} associated to virtual gluons. Gluon has a bosonic spin
($s=1$)
like photons,thus,in this case,a direct classical analogy to magnetization $M$ due to fermions ($s=\pm 1/2$) leads us to think about a kind of
``color
magnetization $M_a$" for QCD-vacuum as being a ``color paramagnetic medium" in the presence of a ``color magnetic field $H_a$". Following such
an analogy to QCD-vacuum, we can write:
\begin{equation}
M_a=\chi_{eff.}H_a,
\end{equation}
where $a$ is just a color mode that we select for convenience, being $\chi_{eff.}=\chi(u)$ the ``effective paramagnetic susceptibility" for QCD- vacuum,
with dependence on the energy scale $u$.

 In (8), paramagnetism manifests itself through the minus sign in front of the right-hand
side, which shows, in analogy to QCD-vacuum, that vacuum energy in QCD is decreased in the presence of a color magnetic field\cite{18}. So (8) can be
written in the following way for representation of the ``color paramagnetic energy" of QCD-vacuum, namely:

\begin{equation}
E=E_{vac,QCD} = E_{color-paramagnetic}= -\frac{1}{2} 4\pi M_aH_aV = -\frac{1}{2} 4\pi \chi(u) H_a^2 V,
\end{equation}
where $V$ is a kind of coherence volume inside which color fields are greatly correlated, in analogy to correlated spin effect\cite{18}.

The behavior of the increasing function $\chi_{eff.}=\chi(u)$ will be shown in the next section.

\section{An effective Landau's Hamiltonian as a theory for representation of color paramagnetism}

The interesting classical analogy between QCD-vacuum at a certain energy scale $u$ of investigation and a paramagnetic medium with magnetization $M$ in an
external magnetic field $H$ motivates us to introduce an effective Landau's Hamiltonian for representation of vacuum inside the hadronic matter as a
paramagnetic medium in the presence of a color magnetic field $H_a$. This simple model will be presented in this section.

It is well-known that a cloud of virtual gluons emerges in QCD-vacuum at higher energies $u$, leading to the ``color paramagnetism"
 (antiscreening),whereas,
on the other hand, a cloud of virtual electron-positron appears in QED-vacuum at higher energies, leading to vacuum polarization. We have a ``dielectric
vacuum" (screening) for QED.

The cloud of virtual gluons of the QCD-vacuum are quanta of the color field induced by the probe used to investigate the structure of the hadronic matter
and depends on its energy scale $u$. But it depends also on the proper color magnetic field $H_a$ that already exists inside the hadronic matter under
investigation. Therefore, such a color field could be thought as
being directly related to the color magnetization $M_a$ and also to the color magnetic field $H_a$ since we have the relation $M_a=\chi(u)H_a$. So now
let us think about such a color field as being a general scalar field $\Phi_a$, namely:

\begin{equation}
\Phi_a=\Phi_a(r)=\Phi_a[M_a(r),H_a(r)]=\Phi_a[\chi(u),H_a(r)],
\end{equation}
where $M_a(r)=\chi(u)H_a(r)$. As the effective susceptibility $\chi_{eff.}=\chi(u)$ and the color magnetic field $H_a$ are independent parameters, let us
use for convenience the scalar color field $\Phi_a(r)$ in the form $\Phi_a(r)=\Phi_a[\chi(u),H_a(r)]$. Here we now think that the color magnetic field
$H_a$ and color magnetization $M_a$ have dependence on $r$-coordinate inside the ``color paramagnetic medium" represented by the hadronic matter.

Due to the classical analogy to paramagnetism, let us introduce now the following ``effective Landau Hamiltonian" for ``color paramagnetism", namely:

\begin{equation}
F=\int[~[\nabla\Phi_a(\chi,H_a(r))]^2 + R(L)\Phi_a(\chi,H_a(r))^2 + K(L)\Phi_a(\chi,H_a(r))^4~]dr,
\end{equation}
where, in this case, the coefficients $R(L)$ and $K(L)$ do not depend on temperature since there is not any phase transition in such effective model.

The integration (12) extends over $d$-dimensional volume. Thompson's approach has three assumptions (see ref.[3]). As we are not interested in
phase transition, we will use only the two first ones, namely:

(A) When the integral in (12) is taken over the scale volume $L^d$ in $d$-dimensions, the three terms separately in (12) are all of order unity.

(B) In the specific case of (12), we just have to consider the parameter $K(L)$ to be finite in the limit $L\rightarrow\infty$. This leads us to consider a mean field regime above a certain critical dimension $d_c$, where the coefficient $K$ remains contant. In Landau's theory, $d_c=4$~(ref.[3]).

By applying the assumption (A) in the first term of (12), we write:

\begin{equation}
\int_{L^d} (\nabla\Phi_a)^2 dr\sim 1,
\end{equation}
where the parameter $L$ forms the basis of our dimensional argument and may be thought as a wavelength cut-off, so that the mean value
$\overline\Phi_a^2$ behaves as

\begin{equation}
\overline{\Phi_a^2}\sim L^{2-d}
\end{equation}
For the second term in (12) we have

\begin{equation}
\int_{L^d} R(L)\Phi_a^2 dr\sim R(L)\overline{\Phi_a^2} L^d\sim 1
\end{equation}
By introducing (14) into (15), we obtain

\begin{equation}
R(L)L^2\sim 1.
\end{equation}

For the third term in (12) we have

\begin{equation}
\int_{L^d} K(L)\Phi_a^4 dr\sim K(L)\overline{\Phi_a^4} L^d\sim 1,
\end{equation}
such that from (14) and assumption (B) we obtain from (17)
\begin{equation}
 K(L)=\left\{
\begin{array}{ll}
 L^{d-4}:&\mbox{$d\leq 4$},\\\\
 1:&\mbox{$d\geq 4$}
 \end{array}
\right.
\end{equation}

From (18), we observe that $d=4$ is a special dimension (an upper critical dimension) above which we have a mean field behavior\cite{6}, that is to
say, the coupling parameter $K$ does not depend on scale $L$, being a constant parameter. In other words, below $d=4$, fluctuations are very important
for the problem, deviating from mean field behavior, and above $d=4$,``mean field" description\cite{6} is a good description for the problem. So
$d=4$, which coincides with the space-time dimensions, corresponds exactly to a kind of border-line dimension to represent QCD-vacuum as a Lorentz
invariant theory and also its classical analogy with paramagnetic media. Therefore, we must improve our approximations in order to ``see" the
logarithmic dependence on scale $L$ (ref.[17]) of the coupling $K(L)$ just in $d=d_c=4$, or equivalently on the energy scale $u=L^{-1}$. Similar
situation has also occurred when we treated diffusion limited chemical reactions through Thompson's approach\cite{4}~\cite{6}~\cite{11}~\cite{12}
~\cite{13}, displaying universal logarithmic behavior on ``upper critical dimensions" for ``coupling constants" of those different models. Following
that improvement technique to ``see" such logarithmic behavior, let us improve the calculation of (17) by taking the quantity $\Phi_a^4$ inside the
integral (17), firstly by starting from the same scale form as that evaluated in (14), but now displaying a dependence on the $r$-variable of scale.
So by taking inside the integral (17) the quantity $[\Phi_a]_r^4=([\Phi_a]_r^2)^2= r^{4-2d}$ and also the $d$-dimensional volume of integration in the
form $d^dr=r^{d-1}dr$, we have

\begin{equation}
\int K(r) r^{4-2d}r^{d-1}dr=\int K(r) r^{3-d}dr\sim 1,
\end{equation}
where, only for $d=4$, exactly on the boder-line of mean field regime where $K$ is practically constant, we are able to see now the refinement of the
logarithmic dependence on length scale for $K$, that is, $\int K r^{-1}dr\sim 1$, which implies $K\sim [ln(r)]^{-1}$. So now, if we perform such integration between the limits of scales $L$ and $L_0$, by considering here $L_0$ an upper cut-off of length, we write:

\begin{equation}
\int_{L}^{L_0} K r^{-1} dr\sim 1,
\end{equation}
from where we obtain:

\begin{equation}
K=K(L)\sim\frac{1}{ln(\frac{L_0}{L})}\sim K(u)\sim\frac{1}{ln(\frac{u}{u_0})},
\end{equation}
where the energy scales are $u=1/L$ and $u_0=1/L_0$, being $u>u_0$, since $u_0$ is a lower cut-off in the scale of energy,i.e.,it is a infrared limit.

 As we have obtained the logarithmic behavior of the coupling parameter $K$ just at $d=d_c=4$ for such a paramagnetic medium, we can do a direct
 analogy
with $QCD_4$-vacuum by obtaining now the ``color scalar field amplitude" $\Phi_a^2$ in a direct analogy to the equilibrium magnetization of the Landau
picture, namely $M^2=-\tau r(L)/u(L)$~(ref.[3]). However, since we do not have any spontaneous breakdown of symmetry, we just consider the simple
coefficient $R$ instead of $\tau r(L)$~(ref.[3]). Thus we simply obtain:

\begin{equation}
\Phi_a^2 =-\frac{R}{K}
\end{equation}

As we are interested only in the behavior of $\Phi_a^2$ on the border-line at $d=d_c=4$ associated to our space-time, we introduce (21) into (22) and
so we have:

\begin{equation}
\Phi_a^2(r)=-c_1R(r)~ln(\frac{u}{u_0}),
\end{equation}
where $c_1>0$ is a positive proportionality constant.

We can associate the amplitude of scalar field $\Phi_a^2(r)$ to a negative energy density $\rho(r)$ of a ``color paramagnetic medium" ($QCD_4$-vacuum)
in analogy to that negative energy density of a paramagnetic medium, namely $-\frac{1}{2}4\pi\chi H^2$ (ref.[18]). However, we must consider an
``effective susceptibility" $\chi_{eff.}=\chi(u)$ to represent QCD-vacuum and also a ``color magnetic field" $H_a(r)$. So such analogy leads
us to write:

\begin{equation}
\rho_{vac,QCD}(r)=\Phi_a^2(r)=-c_1R(r)~ln(\frac{u}{u_0})\equiv-\frac{1}{2}4\pi\chi(u)H_a(r)^2,
\end{equation}
from where we can firstly extract $c_1\equiv 2\pi$ and $R(r)\equiv H_a^2(r)$ and so we can rewrite (24) as follows:

\begin{equation}
\rho_{vac,QCD}(r)=\Phi_a^2(r)=-c_1R(r)~ln(\frac{u}{u_0})\equiv -\frac{1}{2}4\pi H_a(r)^2~ln(\frac{u}{u_0}).
\end{equation}

By comparing the right side of (25) with the right side of (24), we can also extract the ``effective susceptibility", as follows:

\begin{equation}
\chi_{eff.}=\chi(u)=ln(\frac{u}{u_0}).
\end{equation}

From (26), it is interesting to observe that the effective susceptibility of $QCD_4$-vacuum increases logarithmicaly with energy scale $u$. We can also
observe that, from (21) and (26), the parameter $K$ is $K(u)\sim\chi(u)^{-1}$, which allows us
to interpret such parameter as being related to a ``strength" of coupling $\alpha_S$ between quarks. So we have $\alpha_S\sim K$. That is because, when
$u\rightarrow u_0$ in infrared limit, this implies in $\chi(u_0)\rightarrow 0$ (very weak ``paramagnetism"), which leads to $\alpha_S(u_0)\sim
K(u_0)\rightarrow\infty$ (a much more strong coupling), that is, we have a highly confined regime of quarks in lower energies. On the other hand, when
$u\rightarrow\infty$ in ultraviolet regime, this implies $\chi(u)\rightarrow\infty$ (``color paramagnetism" become much more evident), which leads to
$\alpha_S(u)\sim K(u)\rightarrow 0$ (a very weak coupling between quarks), that is to say, we have the well-known asymptotic freedom for higher energies.

For sake of simplicity, if we take the color magnetic field $H_a$ practically uniform in (25), that is also to say, an uniform energy density
$\rho_{vac,QCD}$ or $\Phi_a^2$, and by considering a coherence volume $V$, we simply obtain the``color paramagnetic energy" $E$ as that given in
(10), being $\chi_{eff.}$ now given in (26). So we finally write:

\begin{equation}
 E_{vac,QCD}=\rho_{vac,QCD}V= \Phi_a^2 V= -\frac{1}{2}4\pi H_a^2~ln(\frac{u}{u_0})V.
\end{equation}

The ``effective magnetic permeability" $\mu(u)=1+4\pi\chi(u)$ can be obtained by considering (26), namely:

\begin{equation}
\mu(u)=1+4\pi~ln(\frac{u}{u_0})
\end{equation}

In order to obtain the ``effective electric permissivity" or the dieletric constant $\epsilon(u)$ of QCD-vacuum, now we must guarantee the Lorentz
invariance by considering the relation (6) ($\mu\epsilon=1$).  So doing that and considering (28), we find

\begin{equation}
\epsilon(u)=\frac{1}{1+4\pi~ln(\frac{u}{u_0})},
\end{equation}
being $u\geq u_0$. We have $\mu=\mu_0=1$ for $u=u_0$. 

In QCD, we have an antiscreening such that the effective interaction between strong charges for higher energies is $Q_{eff.S}^2=\epsilon q_S^2$, with $\epsilon<1$, that is $Q_{eff.S}<q_S$. As the strong interaction is directly related to the strong coupling $\alpha_S$, we can also write it in the
form: $\alpha_{S}=\epsilon~\alpha_{0S}$. So by considering (29), finally we can also write it as follows:

\begin{equation}
\frac{\alpha_S}{\alpha_{0S}}=\frac{1}{1+4\pi~ln(\frac{u}{u_0})},
\end{equation}
where we fix $\alpha_{0S}$ to be a large value, but finite for lower energies. So (30) reveals to us the asymptotic freedom behavior for $QCD_4$ at
higher energies because, if we fix $u_0$ and consider $u\rightarrow\infty$, the ratio $\frac{\alpha_S}{\alpha_{0S}}\rightarrow 0$, which means that
the strong coupling decreases when the energy scale increases. However, actually, here only the bosonic contribution of gluons in QCD-vacuum was evaluated
for dieletric constant. In reality, there is a competition between the effects of bosonic (antiscreening) and fermionic (screening) contributions, where
the first one prevails. This subject will be treated in the next section.

\section{Contribution of quantum fluctuations for energy}

 Quantum fluctuations lead to an interaction energy ($\Delta mc^2$) as an increment in the field energy and with logarithmic behavior on
length or energy scale, and we can represent both fermionic and bosonic contributions of energy density $u_{F}$ and $u_{B}$ (see (2)) in the following
conpact form:

\begin{equation}
u_T= u_{qF} + u_{qB}=\frac{1}{2}\frac{1}{4\pi}\alpha\hbar c \frac{1}{r^4}(\frac{r}{\overline\lambda_c}),
\end{equation}
where $u_T$ is the total contribution of quantum fluctuations for energy density, such that there is certain superior cut-off wavelength
$\overline\lambda_c$,below which ($r<\overline\lambda_c$) we have quantum behavior of energy density $u_T$ on scale-$r$, that is,
$u_T=u_q\propto 1/r^3$ (see ref.[3]), and equal or above which ($r\geq\overline\lambda_c$) we recover the well-known classical behavior of $u_T$, that is,
$u_T=u_{cl.}\propto 1/r^4$. To be more accurate, we rewrite a general form of $u_T$ in two regimes, namely:

\begin{equation}
 u_T=\left\{
\begin{array}{ll}
  u_{q}=u_{qF}+u_{qB}=\frac{1}{2}\frac{1}{4\pi}\frac{\alpha\hbar c}{\overline\lambda_c r^3}: &\mbox{$r\leq\overline\lambda_{c0}$}\\\\
  u_{cl}=\frac{1}{2}\frac{1}{4\pi}\frac{\alpha\hbar c}{r^4}: &\mbox{$r\geq\overline\lambda_{c0}$},
 \end{array}
\right.
\end{equation}
where $\overline\lambda_{c0}=\hbar/m_0c$ is a sharp cut-off wave-length. As the quantum regime also presents the bosonic contributions $u_B$ which
leads to antiscreening in QCD, the mass $m_0$ must be considered as a dynamical variable which exhibits fluctuations depending on the energy scale.

We are interested only in quantum regime for energy density $u_q$ of the field ($r<\overline\lambda_{c0}$). So we want to obtain the interaction
energy $\Delta E=\Delta m c^2$ in a certain coherence volume $V$. Then let us think in a spherical volume $V$ and therefore we have the interaction energy in the differential form, namely:

\begin{equation}
dE=dmc^2=u_q 4\pi r^2dr=(\frac{1}{2}\frac{1}{4\pi}\frac{\alpha\hbar c}{\overline\lambda_c r^3})4\pi r^2dr.
\end{equation}
We also can write (33) in the following way:

\begin{equation}
dE=dmc^2=\frac{1}{2}\frac{\alpha\hbar c}{\overline\lambda_c}(\frac{dr}{r})=-\frac{1}{2}\frac{\alpha\hbar c}{\overline\lambda_c}(\frac{du}{u}),
\end{equation}
where we have considered the energy scale $u$ such that $r=u^{-1}$, being $dr/r=udr=-du/u$. So by performing the integration of (34), we write:

\begin{equation}
\Delta mc^2=-\frac{1}{2}\int_{u_0}^{u}\frac{\alpha\hbar c}{\overline\lambda_c}(\frac{du}{u})=
-\frac{1}{2}\int_{u_0}^{u} \alpha mc^2(\frac{du}{u}),
\end{equation}
where $\overline\lambda_c=\hbar/mc$, and $u>u_0$.
Due to fluctuations, it is natural to think that the coupling $\alpha$ and the mass $m$ vary rapidly with the energy scale $u$, such that we must
take them off the integral (35) as avarage values on scale $u$, namely:

\begin{equation}
\Delta m = -\frac{1}{2}{\overline\alpha}~{\overline m}\int_{u_0}^{u}\frac{du}{u},
\end{equation}
where we define $(\alpha m)_{av}= {\overline\alpha}~{\overline m}$.

On the other hand,it is important to perceive that the increment on mass $\Delta m=(\int_Vu_qdV)/c^2$ due to interactions in such quantum regme is
directly proportional to the increment on charge or coupling ($\Delta\alpha$) since both the increments present logarithmic behavior on scale. Such
reasoning was used before in $QED$ ( see ref.[17]) and so extending it to our problem, let us write:

\begin{equation}
\frac{\Delta\alpha}{\overline\alpha}=\frac{\Delta m}{\overline m}
\end{equation}

By introducing (36) for $\Delta m$ into (37) and performing the calculations, we obtain:

\begin{equation}
\alpha=\alpha_0 - \frac{1}{2} (\overline\alpha)^2~ln(\frac{u}{u_0}),
\end{equation}
being $\Delta\alpha=\alpha-\alpha_0$.

Now let us write $(\overline\alpha)^2$ in the following way:
\begin{equation}
(\overline\alpha)^2=\alpha\alpha_{ref},
\end{equation}
where $\alpha_{ref}$ is a certain reference coupling to be duly interpreted.

The equations (36),(37) and (39) define the variable parameters $\overline\alpha$, $\overline m$ and $\alpha_{ref}$. Only the parameters $\alpha$,
$m$, $\Delta\alpha$ and $\Delta m$ are always real quantities since they are physical parameters. 

By substituting (39) in (38), we obtain:
\begin{equation}
\alpha=\alpha(u)=\frac{\alpha_0}{1+\frac{\alpha_{ref}}{2}~ln(\frac{u}{u_0})}.
\end{equation}

From the general result (40), we can observe that the sign of $\alpha_{ref}$ can change by controlling the predominance of antiscreening or screening.
In the case of $\alpha_{ref}>0$, then we have $\alpha\rightarrow 0$ for $u\rightarrow\infty$, which leads to the asymptotic freedom behavior of QCD
connected to antiscreening. On the other hand, if $\alpha_{ref}<0$, we have the well-known Landau singularity\cite{17}, namely a finite value of the
energy scale $u_L$ such that $\alpha(u_L)\rightarrow\infty$. This case of increasing of $\alpha$ with the increasing of $u$ is the behavior of QED-
coupling $\alpha$ associated to screening. Such opposing case ($\alpha_{ref}<0$) leads to an imaginary value for $\overline\alpha$ since $\alpha$
must be a real and positive value (see (39)) 

The result (40) also implies the following differential equation:

\begin{equation}
u\frac{d\alpha}{du}= -(\frac{\alpha_{ref}}{\alpha_0})\alpha^2,
\end{equation}
that is, by performing the integration of (41) above in the limits $u_0$ and $u$ and their respective couplings $\alpha(u_0)$ and $\alpha(u)$, we obtain (40).

In respect to sign of the term $\alpha_{ref}/\alpha_0$ which appears in the right side of the differential equation (41), we must interpret $\alpha_{ref}$
as being a general parameter that includes an effective result of a competition between fermionic and bosonic contributions, namely the final result
of the competition between screening and antiscreening effect. So let us write:

\begin{equation}
\frac{\alpha_{ref}}{\alpha_0}=\frac{N_B - N_F}{Q},
\end{equation}
where $Q$ is simply a number to be determined by using boundary condition. $N_B$ is interpreted as a number of total bosonic species, contributing for
antiscreening, namely a kind of total number of ``normal modes" for bosonic field to be defined. $N_F$ is a number of fermionic species, contributing for screening, which will be also defined.

  At higher energy scales, whereas a cloud of virtual pairs of particle $e^{-}~e^{+}$ is induced in QED-vacuum as a dieletric meduim, similarly a
cloud of virtual gluons $g~\overline g$ is induced in QCD-vacuum as a paramagnetic medium\cite{18}. Following now such semiclassical analogies, let us
look for an alternative way to count $N_B$ and $N_F$.

  In QED, each pair $e^{-}e^{+}$  plays the role of a electric dipole in ``dieletric vacuum" ($\epsilon>1$), which has only one flavor, namely
$n_{F\overline F}=1$, where $F$ represents fermion ($e^{-}$) and $\overline F$ represents anti-fermion ($e^{+}$). In general, as each quark flavor
has spin degenerecency $\pm 1/2$ due to its fermionic property, so we count the total $N_F$ as being

\begin{equation}
N_F=2n_F=2n_{F\overline F},
\end{equation}
where $n_{F\overline F}$ or simply $n_F$ represents an effective number of flavors $F$. $\overline F$ is already included into this counting because each
pair $F\overline F$ forms one ``flavor electric dipole". The multiplying factor $2$ comes from the degenerecency of spin $2s+1$ for $s=1/2$.

  On the other hand, a semiclassical anology between QCD-vacuum (asymptotic freedom) as spin effect\cite{18} due to virtual gluons which play the role
 of microscopic magnetic dipoles leads us to think $N_B$ as being

\begin{equation}
N_B=2(n_g + n_{\overline g}),
\end{equation}
where, although each virtual pair $g\overline g$ emerges in QCD-vacuum, $g$ and $\overline g$ are treated separately as being each one a magnetic dipole, that is, each virtual gluon in QCD-vacuum works effectively like a spin effect of a paramagnetic medium for this semiclassical analogy. $n_g$ is associated to a number of gluon states. Here, the multiplying factor $2$ comes from the degenerescency due to two possible polarizations of gluon, which works like a photon in the sense that it is a massless particle.

Introducing now (44) and (43) into (42) and after into (41), we obtain

\begin{equation}
u\frac{d\alpha}{du}= - (\frac{2(n_g + n_{\overline g}) - 2n_{F\overline F}}{Q})\alpha^2,
\end{equation}
where $Q$ must be obtained by using boundary condition.

The differential equation (45) is an alternative way to get a unified vision between QCD and QED. It must coincide with the special case of
QED-$\beta$ function when both of numbers $n_g=n_{\overline g}=0$ (no gluon state) and also the number $n_{F\overline F}=1$, which means only one
flavor. So doing that and comparing the result with that well-known QED-$\beta$ function\cite{17} evaluated at one loop level, we have

\begin{equation}
 \left[u\frac{d\alpha}{du}\right]_{QED}=\frac{2}{Q}\alpha^2\equiv\frac{2}{3\pi}\alpha^2,
\end{equation}
from where we extract $Q=3\pi$. So returning to (45) and admitting now the case of QCD, where $n_g=n_{\overline g}=8$
(the well-known 8 states of gluon comming from SU(3) symmetry), we finally obtain

\begin{equation}
 \left[u\frac{d\alpha}{du}\right]_{QCD}=-(\frac{32 - 2n_F}{3\pi})\alpha^2
\end{equation}

This result (eq. (47)) must be compared with the well-known $\beta$-function for QCD\cite{18}~\cite{27}~\cite{28} when evaluated at one loop level,
namely:

\begin{equation}
 \left[u\frac{d\alpha}{du}\right]_{QCD}=-(\frac{33 - 2n_F}{3\pi})\alpha^2
\end{equation}

It would be worth to make some comparison of the results of equation (47) with some experimental evaluation of the strong coupling as a function
of momentum (energy) of the probe. In a plot of reference (19) it is possible to get an estimate for $\alpha_0$, namely:

\begin{equation}
\alpha_0=\alpha(u_0=1Gev)\cong 0.43
\end{equation}

Taking in account (41) and (47), we can write

\begin{equation}
\alpha_{ref}=\alpha_0\frac{20}{3\pi}\cong 0.91
\end{equation}

In obtainning $\alpha_{ref}$ in eq. (50), we have considered $n_F=6$ as the number of quark flavors.

By using eq.(40) and the fact that $\alpha_{ref}\equiv\alpha(u=u_{ref})$, $u_{ref}$ can be determined. Taking in account the previous results,
we get

\begin{equation}
u_{ref}\cong 383 Mev.
\end{equation}

This is slightly greater than the quark constituent mass of the nucleon.

On the other hand, if we consider $5$ as the number of quarks flavors, we obtain $\alpha_{ref}(n_F=5)\cong 1$ and $u_{ref}(n_F=5)\cong 320$ Mev. This
value is close to quark constituent mass of the nucleon if we consider that each valence quark carries out one third of the nucleon mass.\\\\

{\noindent\bf  Acknowledgedments}

 Dr. Cl\'audio Nassif specially acknowledge FAPERJ by the financial help.


\begin{thebibliography}{00}
\bibitem{1}   K. G. Wilson, Rev. Mod. Phys. {\bf 55}, 583 (1983).

\bibitem{2}   K. G. Wilson, Physica {\bf 73}, 119 (1974).

\bibitem{3}   C. J. Thompson, J. Phys. {\bf A9}, L25 (1976).

\bibitem{4}   P. R. Silva, Phys. Stat. Sol. {\bf B179}, K5 (1993).

\bibitem{5}   L. Peliti, J. Phys. {\bf A19}, L365 (1986).

\bibitem{6}   C. Nassif and P. R. Silva, Mod. Phys. Lett. {\bf B13}, 829 (1999).

\bibitem{7}   A. Aharony, Y. Imry and S.-K. Ma., Phys. Rev. Lett. {\bf 37},1364   (1976).

\bibitem{8}    P. R. Silva, Phys. Stat. sol. {\bf B165}, K79 (1991).

\bibitem{9}   P. R. Silva, Phys. Stat. sol. {\bf B174}, 497 (1992).

\bibitem{10}   P. R. Silva, Phys. Stat. sol. {\bf B179}, K99 (1993).

\bibitem{11}  C. Nassif and P. R. Silva, Mod.Phys.Lett.{\bf B15},33 (2001).

\bibitem{12}  C. Nassif and P. R. Silva, Mod.Phys.Lett.{\bf B15}No 26,1205 (2001).

\bibitem{13}  C. Nassif and P. R. Silva, Mod.Phys.Lett.{\bf B16}, 601 (2002).

\bibitem{14}  C. Nassif and P. R. Silva, Physica{\bf A} 334, 335-342  (2004). 

\bibitem{15}  P. R. Silva, Int. Jour. Mod. Phys.{\bf A} 12 (7),1373-1384 (1997).

\bibitem{16}  C. Nassif and P. R. Silva, Int. Jour. Mod. Phys. B {\bf 17}, 26, 4645 (2003).

\bibitem{17}  C. Nassif and P. R. Silva, Int. Jour. Mod. Phys. A {\bf 21}, 18, 3809 (2006) 

\bibitem{18}  N. K. Nielsen, Am. J. Phys.{\bf 49}, N.12, 1171-1178 (1981) 

\bibitem{19}  F. Wilczek, Phys. Today {\bf August}, 22-28 (2000). 

\bibitem{20} For a wide-ranging survey, see, H. Kastrup, P. Zerwas, eds., $QCD$
20 Years Later, World Scientific, Singapore (1993).
 
\bibitem{21} D. J. Gross and F. Wilczek, Phys. Rev. Lett.{\bf 30}, 1343 (1973). 

\bibitem{22} H. D. Politzer, Phys. Rev. Lett.{\bf 30}, 1346 (1973). 

\bibitem{23}  Laurent Nottale, in: Fractal Space-time and Microphysics, Ch.6, p.203, World Scientific Publishing Co. Pte. Ltd, Singapore, New Jersey, London, Hong Kong (1993).

\bibitem{24} S. Weinberg, The Quantum Theory of Fields, Vol1, p.496, Cambridge University Press, (USA) (1996).

\bibitem{25} V. F. Weisskopf, Phys. Rev. {\bf 56}, 72 (1939); Phys. Today {\bf November}, 69-85 (1981).

\bibitem{26} C. G. Callan, R. F. Dashen and D. J. Gross, Phys. Rev. D {\bf 19}, 1826 (1979).

\bibitem{27} K. Moriyasu, An Elementary Primer For Gauge Theory, World Scientific, Singapore (1983).

\bibitem{28} F. Halzen and A. D. Martin, Quarks and Leptons: An Introductory Course in Modern Particle Physics, Wiley, N. York (1984).

\end{thebibliography}
\end{document}